\documentclass[12pt]{article}

\usepackage[T1]{fontenc}
\usepackage{helvet}

\usepackage[letterpaper,margin=1in]{geometry}
\usepackage{setspace}
\usepackage{ragged2e}
\usepackage[super,sort,comma]{natbib}
\usepackage{amsmath}
\usepackage{graphicx}
\usepackage{float}
\usepackage{needspace}
\usepackage{booktabs}
\usepackage{caption}
\usepackage[section]{placeins}
\usepackage[hidelinks]{hyperref}
\hypersetup{pdfauthor={Xianghuan He, Yao Hao, Changyu Sun, Andy N. Wiles, Gregory B. Biedermann, Brian W. Pogue, Rongxiao Zhang},pdftitle={Emission-associated energy deposition and Cherenkov--dose proportionality in megavoltage photon beams},pdfsubject={},pdfkeywords={}}
\usepackage[all]{hypcap}
\makeatletter
\renewcommand\@biblabel[1]{#1.}
\makeatother
\renewenvironment{abstract}{\section*{Abstract}\small\noindent}{\par}

\begin{document}
\RaggedRight
\setlength{\emergencystretch}{3em}
\setlength{\parindent}{2em}
\setlength{\parskip}{0pt}
\pagestyle{plain}
\pagenumbering{arabic}

\begin{center}
{\Large\bfseries Emission-associated energy deposition and
Cherenkov--dose proportionality in megavoltage photon beams\par}

\smallskip
{\normalsize
Xianghuan He\textsuperscript{1}, Yao Hao\textsuperscript{2},
Changyu Sun\textsuperscript{1,3}, Andy N. Wiles\textsuperscript{4},\\[3pt]
Gregory B. Biedermann\textsuperscript{4}, Brian W. Pogue\textsuperscript{5},
Rongxiao Zhang\textsuperscript{4,*}\par}
\smallskip
{\footnotesize
\textsuperscript{1}Department of Chemical and Biomedical Engineering, College of Engineering,
University of Missouri, Columbia, Missouri, United States.\par
\textsuperscript{2}Division of Medical Physics, Department of Radiation Oncology,
Washington University School of Medicine, St. Louis, Missouri, United States.\par
\textsuperscript{3}Department of Radiology, School of Medicine,
University of Missouri, Columbia, Missouri, United States.\par
\textsuperscript{4}Department of Radiation Oncology, School of Medicine,
University of Missouri, Columbia, Missouri, United States.\par
\textsuperscript{5}Thayer School of Engineering, Dartmouth College,
Hanover, New Hampshire, United States.\par\smallskip
\textsuperscript{*}Corresponding author: Rongxiao Zhang,
\href{mailto:rongxiao.zhang@health.missouri.edu}{rongxiao.zhang@health.missouri.edu}\par}
\end{center}

\begin{abstract}
\noindent\textbf{Background:} Quantitative conversion from Cherenkov light to absorbed dose depends on beam spectrum, field size, depth, and material properties. Charged particles emit Cherenkov photons only when their energy exceeds the emission threshold. Accurate dose estimation from Cherenkov light requires a reliable conversion relationship.\par

\noindent\textbf{Purpose:} To distinguish dose contributions with and without accompanying Cherenkov production, examine their spatial relationships with local Cherenkov production, and assess their implications for calibration.\par

\noindent\textbf{Methods:} A Geant4-based digital linac framework generated paired Monte Carlo dose and local Cherenkov distributions in a water phantom for flattened  6, 10, and 15 MV and flattening-filter-free  6 and 10 MV beams with 3 $\times$ 3 and 10 $\times$ 10 cm$^{2}$ fields.  Total deposition was separated into emission-tagged and non-emission-tagged components according to whether each step actually produced at least one Cherenkov photon. Component energy fractions were calculated over the phantom. Distributions were independently normalized within a shared reference region and evaluated using voxelwise Cherenkov-to-dose ratios and lateral profiles. Spatial partitioning was used to identify the origins of the ratio structures.\par

\noindent\textbf{Results:} Emission-tagged energy fractions were 37\%--61\%, increasing with beam energy and lower for FFF beams. Across the five beam qualities, component energy fractions differed by less than 1.3 percentage points between field sizes. Emission-tagged Cherenkov-to-dose ratios formed a narrow dominant band near unity across multiple spatial regions, with 98.4\%--99.9\% of the evaluated ratios between 0.95 and 1.05,  whereas total and non-emission-tagged ratios departed from unity with depth and lateral position. Emission-tagged lateral profiles generally showed smaller residual differences.\par

\noindent\textbf{Conclusions:} Emission-tagged deposition closely tracked local Cherenkov production. Departures of the total Cherenkov-to-dose ratio followed the spatial variation of the emission-tagged dose fraction, which shows where calibration needs correction. The classification reflects actual photon production and depends on transport settings; dedicated electron-energy scoring is needed to isolate sub-threshold contributions.\par
\end{abstract}
\clearpage

\section{Introduction}

Reliable conversion of Cherenkov light to absorbed dose is central to quantitative optical dosimetry in radiotherapy.
Although Cherenkov imaging visualizes radiation delivery,\cite{Glaser2014VideoRateCherenkov} quantitative interpretation depends on the relationship between energy deposition and local photon production.
Theoretical and Monte Carlo studies show strong spatial correspondence between absorbed dose and local Cherenkov production, with differences in normalized spatial distributions depending on beam quality, field size, and depth.\cite{Zhang2013SuperficialPhoton,Glaser2014CherenkovDose,Zlateva2019Formalism}
These differences have been attributed mainly to spectral changes, such as beam hardening with depth.\cite{Glaser2014VideoRateCherenkov,Glaser2014CherenkovDose}
Measured optical signals also depend on tissue optical properties and collection geometry.\cite{Zhang2017BeamTissueCherenkov}
This spatial correspondence does not establish a constant local conversion factor.
For calibration, the question is which parts of the deposited energy follow photon production closely and which account for these differences.

Energy deposition and Cherenkov production weight the electron spectrum differently.\cite{Glaser2014CherenkovDose,Zlateva2019Formalism}
To illustrate this dependence, a schematic continuous-slowing-down description, with knock-on electron energy deposited locally, gives
\begingroup
\begin{align}
D(\mathbf r)&\simeq\frac{1}{\rho}\int_0^{E_{\max}}
\Phi(E,\mathbf r)S_{\mathrm{col}}(E)\,\mathrm dE,\label{eq:spectral_weighting}\\
c(\mathbf r)&\simeq\int_{E_{\mathrm c}}^{E_{\max}}
\Phi(E,\mathbf r)Y_{\mathrm C}(E)\,\mathrm dE,\label{eq:cherenkov_spectral_weighting}
\end{align}
\endgroup
where \(D(\mathbf r)\) is the absorbed dose at position \(\mathbf r\), \(E\) is electron kinetic energy, \(\Phi\) is the electron fluence differential in energy, \(S_{\mathrm{col}}\) is the linear collision stopping power, \(\rho\) is the medium density, and \(c(\mathbf r)\) is the number of Cherenkov photons created per unit volume at position \(\mathbf r\).
\(Y_{\mathrm C}\) is the photon yield per unit track length integrated over a fixed optical band; \(E_{\mathrm c}\) and \(E_{\max}\) are the minimum emission threshold in that band (about 0.24~MeV for the refractive indices used here) and the upper electron-energy limit.
If \(\Phi(E,\mathbf r)=A(\mathbf r)f(E)\), with a spatially invariant spectral shape \(f\), both quantities share the same spatial factor, giving an approximately constant \(D/c\) in a homogeneous medium.
Sub-threshold electrons contribute to dose but not to Cherenkov production; changes in spectral shape can therefore alter the conversion.
This motivates separating deposited energy by its association with photon production.
We classify each deposit by whether its transport step produced Cherenkov photons.
Unlike a classification by electron kinetic energy, this ties each deposit directly to the photons counted.

Here, we propose an emission-associated decomposition of deposited energy using an existing Geant4-based digital linac framework.
We compare total dose and its emission-tagged and non-emission-tagged components with local Cherenkov production in a water phantom for flattened 6, 10, and 15~MV and flattening-filter-free 6 and 10~MV beams at two field sizes.
To inform calibration, we examine how each component relates spatially to local Cherenkov production and where departures of the total Cherenkov-to-dose ratio originate.

\section{Materials and Methods}

\subsection{Paired simulation and emission-associated energy deposition}\label{sec:paired_simulation}

Simulations used our Geant4-based digital linac framework with precomputed TrueBeam phase-space particle records as the incident source.\cite{Agostinelli2003Geant4}\textsuperscript{,}\cite{Allison2016Geant4,Constantin2011TrueBeam,Capote2006PhaseSpace}
Particles were transported through the existing downstream jaw geometry into a water phantom with CT-derived geometry.

Particle transport was modeled in Geant4 11.3.2 using the \texttt{FTFP\_BERT} physics list with standard electromagnetic Option 3.
Secondary-particle production used a range cut of 0.01~mm (production thresholds in water of about 14~keV for electrons and positrons and 1~keV for photons).
The default Option 3 ionization step function for electrons and positrons (range fraction 0.2, final range 0.1~mm) was retained.
Cherenkov step limitation used the default values: a maximum mean yield of 100 photons per step and a maximum 10\% relative change in the particle speed, \(\beta\), per step.
Water refractive indices ranged from 1.3435 to 1.3608 over photon energies of 2.034--4.136~eV, covering a wavelength range of approximately 300--610~nm.\cite{Agostinelli2003Geant4,Allison2016Geant4}
Scintillation was disabled;
 the analysis used only Cherenkov photon creation positions.

If at least one Cherenkov photon was produced during a step, the deposited energy was assigned to the emission-tagged component, \(E_{\mathrm{emit}}\); otherwise, it was assigned to the non-emission-tagged component, \(E_{\mathrm{no\,emit}}\).
No electron-energy selection was applied.
Because Geant4 samples the number of photons in each step from a Poisson distribution, above-threshold steps with a small mean yield, mainly near threshold, may produce no photon and are then non-emission-tagged.
Energy carried by secondary electrons produced above the production threshold is scored in their own steps, and local deposits from photon interactions are always non-emission-tagged.
The partition therefore depends on the production threshold and step limits.
When a step crossed scoring-cell boundaries, its deposited energy was apportioned by the path length in each cell while retaining the same tag.
Both components could contribute to the same voxel.
For voxel \(v\), the total deposited energy is
\begingroup
\begin{equation}
E_{\mathrm{total}}(v)=E_{\mathrm{emit}}(v)+E_{\mathrm{no\,emit}}(v),
\label{eq:decomposition}
\end{equation}
\endgroup

Cherenkov creation positions were separately binned onto the same grid to obtain the Cherenkov photon count \(C(v)\).
The corresponding total, emission-tagged, and non-emission-tagged dose quantities are denoted by \(D_{\mathrm{total}}(v)\), \(D_{\mathrm{emit}}(v)\), and \(D_{\mathrm{no\,emit}}(v)\), respectively.

\subsection{Simulation conditions and comparative analysis}

The simulations comprised flattened 6, 10, and 15~MV beams (6X, 10X, and 15X) and flattening-filter-free 6 and 10~MV beams (6FFF and 10FFF), each with fixed jaw-defined \(3\times3\) and \(10\times10\)~cm\(^{2}\) fields, with the MLC disabled.
Ten conditions were simulated.

Field sizes were defined at the 100-cm source-to-isocenter plane, approximately coincident with the water entrance.
Within each beam quality, the two fields replayed the same phase-space source, with more than \(2.1\times10^{9}\) phase-space particle records per simulation and one full-source realization per condition.

The CT-derived phantom spanned \(51.0\times51.0\times51.75\)~cm\(^{3}\), with a native \(512\times512\times207\) grid.
The BODY (external contour) was assigned water at 1~g/cm\(^{3}\), with air outside.
Energy deposition and Cherenkov creation positions were scored on a common \(203\times203\times201\) grid with 2.5~mm isotropic spacing and the same BODY support.

To assess whether the relative contributions of emission-tagged and non-emission-tagged energy deposition varied with field size, their energy fractions were compared between the two field sizes for each beam quality.
For each condition, the raw deposited energies \(E_{\mathrm{emit}}(v)\), \(E_{\mathrm{no\,emit}}(v)\), and \(E_{\mathrm{total}}(v)\) were summed over the complete BODY, and each component sum was divided by the total-energy sum.

Each distribution was divided by its own mean within the same spherical region of interest (ROI) with a radius of 0.5~cm centered on the central axis at the depth of maximum total dose, \(d_{\mathrm{max}}\).
In this study, \(C\) and the three dose quantities denote these independently normalized distributions, with \(D\) indicating the dose quantity being compared.
Voxelwise \(C/D\) ratios were evaluated for all three dose quantities within a common mask of BODY voxels in which total dose was at least 10\% of its maximum and all ratios were finite.
Ratios were plotted against the corresponding normalized dose quantity; \(C/D=1\) denotes equal normalized values.
Two-dimensional histograms used 120 dose bins and 480 ratio bins over \(C/D=0\)--2.4, with common bin edges and a logarithmic voxel-count color scale across all conditions and components.

To identify the spatial origins of features in the ratio distributions, emission-tagged and non-emission-tagged ratio points were mapped back to the original three-dimensional voxel coordinates. Buildup voxels had depth \(d<d_{\mathrm{max}}\). At or beyond \(d_{\mathrm{max}}\), the field interior satisfied \(\max(|X|,|Z|)\leq0.9h(d)\), where \(h(d)\) is the divergence-adjusted geometric field half-width; the remaining evaluated voxels formed the field-edge region. Water--air interface layers were identified separately by depth.

Lateral X profiles were sampled at total-dose \(d_{\mathrm{max}}\) and \(d_{10}\), using the central \(Z=0\) plane for all quantities.
Profiles retained the spherical-region normalization.
Difference curves were calculated as \(C-D\).
\begingroup
\subsection{Sensitivity and statistical uncertainty}\label{sec:sensitivity_methods}

We used the 6X, \(10\times10\)~cm\(^{2}\) condition for both sensitivity checks and statistical uncertainty analysis.
The sensitivity checks separately varied the production cut and the StepFunction parameters for electrons and positrons, with other settings fixed (Table~\ref{tab:sensitivity_uncertainty}).
The raw photon-to-energy ratio \(C/E_{\mathrm{emit}}\) and normalized spatial agreement between Cherenkov production and emission-tagged dose were compared.
Each setting used one full-source realization.

For the uncertainty analysis, we divided the full phase-space file into 23 non-overlapping batches representing equal numbers of source histories.
Compared with the full-file simulation, only source partitioning and random seeds differed, while all other simulation settings and the total source statistics remained unchanged.
Assuming independent source batches, we estimated voxelwise uncertainties for the full-source simulation from the variation across batches (Table~\ref{tab:statistical_uncertainty}).
\par
\endgroup

\Needspace{8\baselineskip}
\section{Results}

\subsection{\texorpdfstring{Component energy fractions}{Component energy fractions}}

Table~\ref{tab:energy_fractions} summarizes the emission-tagged and non-emission-tagged energy fractions for the five beam qualities and two field sizes, together with local ROI fractions and absolute photon-to-energy ratios.
The emission-tagged fraction increased with nominal energy, from 37.1\%--44.8\% at 6~MV to 61.0\%--61.5\% at 15~MV, and was lower for FFF than for flattened beams of the same nominal energy.
Within each beam quality, the emission-tagged fraction was 0.3--1.3 percentage points lower for the \(10\times10\) than for the \(3\times3\)~cm\(^{2}\) field.
This change is small compared with the differences between beam qualities and is consistent with more low-energy scattered radiation in the larger field. Local emission-tagged fractions at \(d_{\mathrm{max}}\) and \(d_{10}\) were higher than the whole-BODY fractions. In the \(d_{\mathrm{max}}\) ROI, absolute \(C/E_{\mathrm{emit}}\) values showed less relative variation across conditions than \(C/E_{\mathrm{total}}\).

\begin{table}[!htbp]
\centering
\caption{Fractions of total deposited energy from steps with and without actual Cherenkov production, and local photon-to-energy ratios, for the ten conditions.
BODY fractions use raw energy summed over the complete BODY, with the same total-energy denominator for both components. Local fractions use raw energy sums within the 0.5-cm-radius ROIs at \(d_{\mathrm{max}}\) and \(d_{10}\). Here, \(d_{\mathrm{max}}\) is the central-axis depth of maximum total dose, and \(d_{10}\) is a depth of 10~cm from the entrance. The \(C/E\) columns give raw photon-count sums divided by total or emission-tagged energy sums in the \(d_{\mathrm{max}}\) ROI, in photons/MeV.
Field sizes are in cm\(^{2}\).
\label{tab:energy_fractions}}
\small
\setlength{\tabcolsep}{3pt}
\begin{tabular}{@{}lrrrrrr@{}}
\toprule
& \multicolumn{2}{c}{BODY fractions} & \multicolumn{2}{c}{Local emit (\%)} & \multicolumn{2}{c}{\shortstack{ROI $C/E$ @ $d_{\mathrm{max}}$\\(photons/MeV)}} \\
\cmidrule(lr){2-3}\cmidrule(lr){4-5}\cmidrule(l){6-7}
Beam / field & \shortstack{Emission-\\tagged (\%)} & \shortstack{Non-emission-\\tagged (\%)} & $d_{\mathrm{max}}$ & $d_{10}$ & Total & Emit \\
\midrule
6X $3 \times 3$ & 44.82 & 55.18 & 50.4 & 53.0 & 97.13 & 192.53 \\
6X $10 \times 10$ & 43.64 & 56.36 & 48.5 & 48.5 & 93.18 & 192.13 \\
\addlinespace[0.2em]
6FFF $3 \times 3$ & 37.65 & 62.35 & 42.0 & 46.2 & 79.23 & 188.74 \\
6FFF $10 \times 10$ & 37.13 & 62.87 & 40.4 & 41.9 & 76.15 & 188.58 \\
\addlinespace[0.2em]
10X $3 \times 3$ & 57.50 & 42.50 & 64.4 & 65.6 & 133.35 & 207.15 \\
10X $10 \times 10$ & 56.23 & 43.77 & 61.9 & 61.6 & 127.68 & 206.26 \\
\addlinespace[0.2em]
10FFF $3 \times 3$ & 51.62 & 48.38 & 55.8 & 59.1 & 112.98 & 202.65 \\
10FFF $10 \times 10$ & 51.30 & 48.70 & 54.3 & 55.5 & 109.85 & 202.42 \\
\addlinespace[0.2em]
15X $3 \times 3$ & 61.45 & 38.55 & 67.9 & 69.2 & 143.06 & 210.82 \\
15X $10 \times 10$ & 60.95 & 39.05 & 65.7 & 65.6 & 138.03 & 210.05 \\
\bottomrule
\end{tabular}
\end{table}

\subsection{Cherenkov-to-dose ratios}

\begin{figure}[!t]
\centering
\makebox[\textwidth][c]{\includegraphics[width=1.04\textwidth]{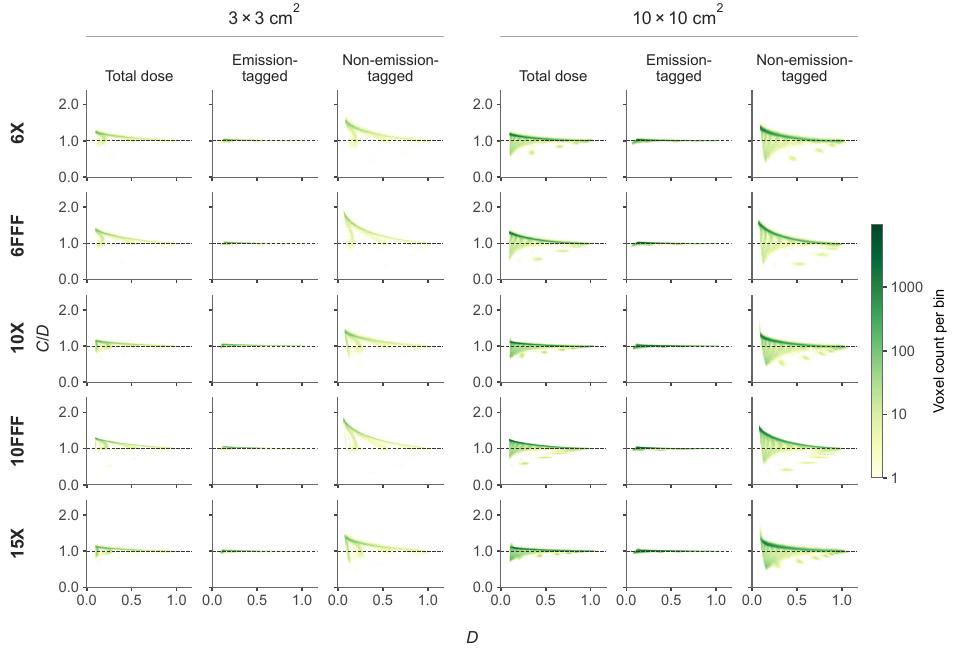}}
\caption{Voxelwise Cherenkov-to-dose ratios for \(3\times3\)~cm\(^{2}\) (left three columns) and \(10\times10\)~cm\(^{2}\) (right three columns) fields.
Rows show 6X, 6FFF, 10X, 10FFF, and 15X.
Within each field, columns correspond to total dose \(D_{\mathrm{total}}\), emission-tagged dose \(D_{\mathrm{emit}}\), and non-emission-tagged dose \(D_{\mathrm{no\,emit}}\).
The horizontal axis shows the indicated dose quantity \(D\), and the vertical axis shows \(C/D\), where \(C\) denotes mass-corrected local Cherenkov photon counts.
Each quantity is normalized to its own mean in the common reference region.
Colors show voxel counts per fixed two-dimensional bin on a logarithmic scale; dashed lines mark \(C/D=1\).
Only voxels with total dose of at least 10\% of its maximum are shown.
\label{fig:ratios}}
\end{figure}

Figure~\ref{fig:ratios} shows the voxelwise Cherenkov-to-dose ratio (\(C/D\)) distributions for total, emission-tagged, and non-emission-tagged dose.
The emission-tagged distribution formed a narrow dominant band near a 1:1 ratio between normalized Cherenkov counts and dose (\(C/D=1\)). Across the ten conditions, 98.4\%--99.9\% of evaluated voxels had \(C/D_{\mathrm{emit}}\) within 5\% of unity, compared with 15.4\%--49.7\% for \(C/D_{\mathrm{total}}\) and 7.2\%--24.3\% for \(C/D_{\mathrm{no\,emit}}\).
The median emission-tagged ratios  for voxels at the entrance surface ranged from 0.932 to 0.998 (Section~\ref{sec:discussion}).
The non-emission-tagged distribution showed three visually distinct regions, particularly evident in the \(10\times10\)~cm\(^{2}\) fields: an upper curved branch declining toward \(C/D=1\) with increasing \(D\), a roughly triangular region below this branch, and isolated clusters at lower ratios.
The total-dose distributions showed the same three features with smaller amplitude.

\FloatBarrier
\subsection{Lateral profiles}

\begin{figure}[!t]
\centering
\makebox[\textwidth][c]{\includegraphics[width=1.10\textwidth]{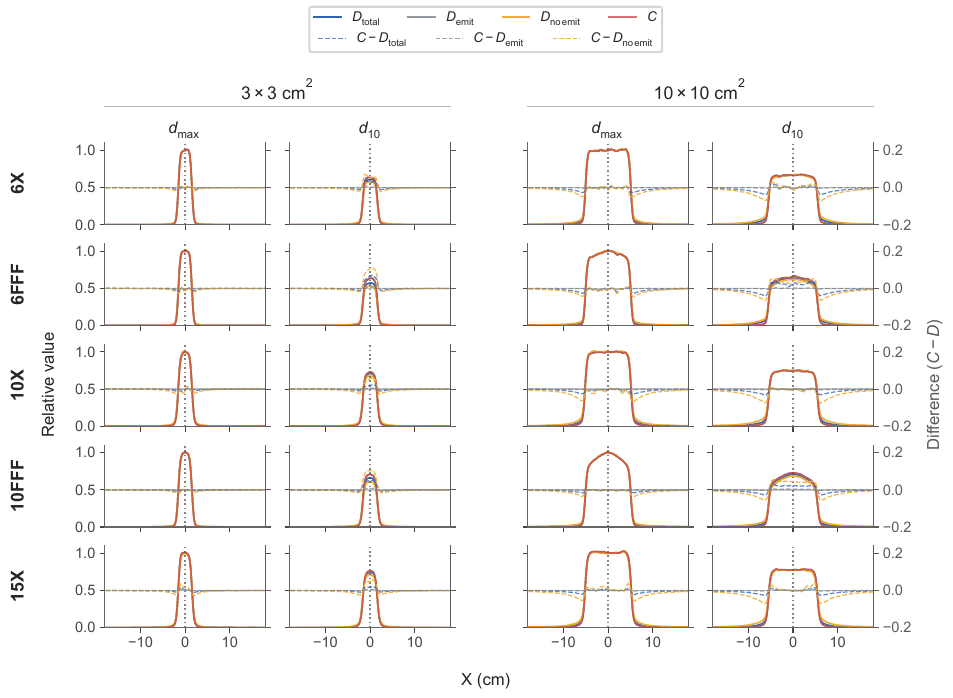}}
\caption{Lateral X profiles at \(Z=0\) for \(3\times3\)~cm\(^{2}\) (left two columns) and \(10\times10\)~cm\(^{2}\) (right two columns) fields.
Rows show 6X, 6FFF, 10X, 10FFF, and 15X; within each field, columns show \(d_{\mathrm{max}}\) and \(d_{10}\). 
Solid curves show total dose (\(D_{\mathrm{total}}\)), emission-tagged dose (\(D_{\mathrm{emit}}\)), non-emission-tagged dose (\(D_{\mathrm{no\,emit}}\)), and local Cherenkov photon counts (\(C\)), each normalized to its own mean in the common reference region.
Dashed curves show \(C-D\) for each dose quantity on the right axes; differences are dimensionless and the horizontal line denotes zero.
\label{fig:lateral_profiles}}
\end{figure}

Figure~\ref{fig:lateral_profiles} compares normalized lateral X profiles of Cherenkov counts and the three dose quantities at two depths: the depth of maximum central-axis total dose (\(d_{\mathrm{max}}\)) and \(d_{10}\).
Within each irradiation condition, the Cherenkov and dose profiles showed similar overall lateral shapes.
At both depths, the emission-tagged dose profiles generally followed the Cherenkov profiles more closely than the total or non-emission-tagged dose profiles.
The emission-tagged \(C-D\) difference curves also generally stayed closer to zero. Across the ten conditions and both depths, the maximum \(|C-D|\) was 0.0093 for emission-tagged dose, compared with 0.0708 for total dose and 0.1128 for non-emission-tagged dose.
The total and non-emission-tagged differences were negative just outside the field edges, where Cherenkov counts fell below dose, and positive on the central axis at \(d_{10}\) in the \(3\times3\)~cm\(^{2}\) fields, most markedly for FFF beams.
Because each profile kept the reference-region normalization, the central-axis differences at \(d_{10}\) reflect the depth dependence of the ratio, not lateral shape.
\FloatBarrier
\begingroup
\subsection{Sensitivity and statistical uncertainty}

Tables~\ref{tab:sensitivity_uncertainty} and \ref{tab:statistical_uncertainty} summarize the three checks for 6X, \(10\times10\)~cm\(^{2}\).
Increasing the production cut from 0.01 to 1~mm raised the BODY emission-tagged energy fraction from 43.64\% to 50.06\%, while total deposited energy changed by less than 0.01\%.
The raw \(C/E_{\mathrm{emit}}\) decreased by 9.29\% and 14.84\% at cuts of 0.1 and 1~mm, respectively, whereas at least 99.22\% of evaluation voxels retained \(R\) within \(1\pm5\%\).
Reducing the StepFunction parameters changed the raw ratio by at most 0.187\%, with normalized spatial agreement nearly unchanged.
At baseline, the reported voxel-level statistical uncertainties ranged from 2.26\% to 3.41\% in the reference region and at 10~cm on the central axis, and from 8.01\% to 9.93\% near the 10\% isodose surface.
\par

\begin{table}[!htbp]
\centering
\captionsetup{font={small,stretch=1.1},labelfont=bf}
\caption{Sensitivity for 6X, \(10\times10\)~cm\(^{2}\).
 Sensitivity relative to baseline settings: production cut 0.01~mm and StepFunction (0.2, 0.1~mm).
Only the indicated setting was changed.
\(C/E_{\mathrm{emit}}\) uses raw BODY sums.
\(R\) is the ratio of mass-corrected Cherenkov counts to emission-tagged dose, each normalized to its own mean in the fixed baseline \(d_{\mathrm{max}}\) ROI.
The same voxels were evaluated across settings.
\label{tab:sensitivity_uncertainty}}
\small
\setstretch{1.0}
\setlength{\tabcolsep}{4pt}
\renewcommand{\arraystretch}{1.12}
\begin{tabular*}{\linewidth}{@{}l@{\extracolsep{\fill}}rrr@{}}
\toprule
\multicolumn{4}{@{}l}{\textbf{Transport-setting sensitivity}} \\
\addlinespace[0.3em]
Setting & \shortstack{\(C/E_{\mathrm{emit}}\)\\(photons/MeV)} & \shortstack{Change\\(\%)} & \shortstack{\(0.95\leq R\leq1.05\)\\voxels (\%)} \\
\midrule
Baseline & 191.921 & --- & 99.36 \\
Production cut: 0.1 mm & 174.097 & \(-9.29\) & 99.33 \\
Production cut: 1 mm & 163.445 & \(-14.84\) & 99.22 \\
StepFunction: (0.1, 0.05 mm) & 191.954 & \(+0.017\) & 99.39 \\
StepFunction: (0.05, 0.025 mm) & 192.279 & \(+0.187\) & 99.37 \\
\bottomrule
\end{tabular*}
\end{table}

\begin{table}[!htbp]
\centering
\captionsetup{font={small,stretch=1.1},labelfont=bf}
\caption{Statistical uncertainty for 6X, \(10\times10\)~cm\(^{2}\).
 Relative standard errors (\(1\sigma\)) at baseline, estimated from 23 source batches.
Multi-voxel rows give medians of individual-voxel uncertainties, not uncertainties of ROI means.
The isodose band spans 9\%--11\% of baseline peak total dose.
\label{tab:statistical_uncertainty}}
\small
\setstretch{1.0}
\setlength{\tabcolsep}{4pt}
\renewcommand{\arraystretch}{1.12}
\begin{tabular*}{\linewidth}{@{}l@{\extracolsep{\fill}}rrr@{}}
\toprule
\multicolumn{4}{@{}l}{\textbf{Relative \(1\sigma\) statistical uncertainty (\%)}} \\
\addlinespace[0.3em]
Location & \(D_{\mathrm{total}}\) & \(D_{\mathrm{emit}}\) & \(C\) \\
\midrule
\(d_{\mathrm{max}}\) reference region (33 voxels) & 2.53 & 3.17 & 3.41 \\
10 cm central axis (one voxel) & 2.26 & 2.75 & 3.05 \\
10\% isodose band (54,725 voxels) & 8.01 & 9.58 & 9.93 \\
\bottomrule
\end{tabular*}
\end{table}
\endgroup

\section{Discussion}\label{sec:discussion}

Every Cherenkov photon is created in an emission-tagged step, and in Geant4 both the mean photon number and the deposited energy of a step scale with its length.
Close agreement between \(C\) and \(D_{\mathrm{emit}}\) is therefore expected in part by construction.
The photon yield per unit emission-tagged energy still varies several-fold with electron energy above threshold, so the narrow band also shows that the above-threshold electron spectrum changed little with  position.

The emission-tagged fraction of dose, \(\eta(v)\), computed from unnormalized doses using Eq.~\eqref{eq:local_emit_fraction}, did vary with position.
For the normalized ratios in Figure~\ref{fig:ratios}, the total-dose ratio is given by Eq.~\eqref{eq:total_ratio_factorization}.
Similarly, the non-emission-tagged ratio is given by Eq.~\eqref{eq:noemit_ratio_factorization}.
\begingroup
\setlength{\jot}{6pt}
\begin{gather}
\eta(v)=\frac{D_{\mathrm{emit}}(v)}{D_{\mathrm{total}}(v)},
\label{eq:local_emit_fraction}\\
\frac{C}{D_{\mathrm{total}}}
=\left(\frac{C}{D_{\mathrm{emit}}}\right)
\frac{\eta(v)}{\eta_{\mathrm{ref}}},
\label{eq:total_ratio_factorization}\\
\frac{C}{D_{\mathrm{no\,emit}}}
=\left(\frac{C}{D_{\mathrm{emit}}}\right)
\frac{\eta(v)/(1-\eta(v))}
{\eta_{\mathrm{ref}}/(1-\eta_{\mathrm{ref}})}.
\label{eq:noemit_ratio_factorization}
\end{gather}
\endgroup
Here, \(\eta_{\mathrm{ref}}\) is the ratio of the reference-region means of the unnormalized \(D_{\mathrm{emit}}\) and \(D_{\mathrm{total}}\).

Where \(C/D_{\mathrm{emit}}\approx1\), departures of the total-dose ratio from unity follow the spatial variation of \(\eta\), and the non-emission-tagged ratios show the same departures amplified by \(1/(1-\eta(v))\).
The structure of the total-dose ratio is therefore explained by where the emission-tagged fraction differs from its value at the calibration point, and that is where Cherenkov-based dose estimates need correction.

The data in Table~\ref{tab:energy_fractions} support the interpretation based on the above factorization. The local fractions distinguish local energy composition from whole-phantom averages and illustrate the spatial variation underlying the \(\eta/\eta_{\mathrm{ref}}\) interpretation. The smaller relative variation of unnormalized \(C/E_{\mathrm{emit}}\) than of \(C/E_{\mathrm{total}}\) across conditions is consistent with changes in the emission-tagged fraction contributing to changes in photon yield per unit total energy.

Figure~\ref{fig:ratios} shows distinct branches and clusters in the non-emission-tagged ratio distribution. These features suggest a dependence on position within the irradiated volume. To examine this possibility, we mapped the ratio points back to their spatial positions, as shown in Figure~\ref{fig:spatial_component_comparison} for the representative 6X \(10\times10\)~cm\(^{2}\) case.
The upper branch mainly corresponds to the field interior at and beyond \(d_{\mathrm{max}}\), while the triangular region corresponds mainly to the field-edge region.
Figure~\ref{fig:spatial_component_comparison}(h) shows that \(\eta(v)/\eta_{\mathrm{ref}}\) generally increases with depth in the field interior and is lower near the field edges, consistent with the spatial variation of the total-dose ratio described by the above factorization.
The rise of the ratio with depth in the field interior is consistent with beam hardening, which increases the share of dose deposited by above-threshold electrons.
Beam hardening was previously identified as the cause of Cherenkov overestimation of dose beyond the depth of maximum dose,\cite{Glaser2014VideoRateCherenkov,Glaser2014CherenkovDose} with larger effects for FFF beams.\cite{Glaser2014CherenkovDose}
Lower ratios at the field edges agree with the softer spectra there, where more of the dose comes from sub-threshold electrons.\cite{Cloutier2022Polarization}
The low-ratio clusters near the entrance correspond to the interface-crossing layer and subsequent buildup layers, where contaminant and low-energy electrons likely deposit dose with little Cherenkov emission before transient charged-particle equilibrium is reached.\cite{Glaser2013Projection}

\begingroup
\setlength{\intextsep}{6pt}
\begin{figure}[H]
\centering
\makebox[\textwidth][c]{\includegraphics[width=1.10\textwidth]{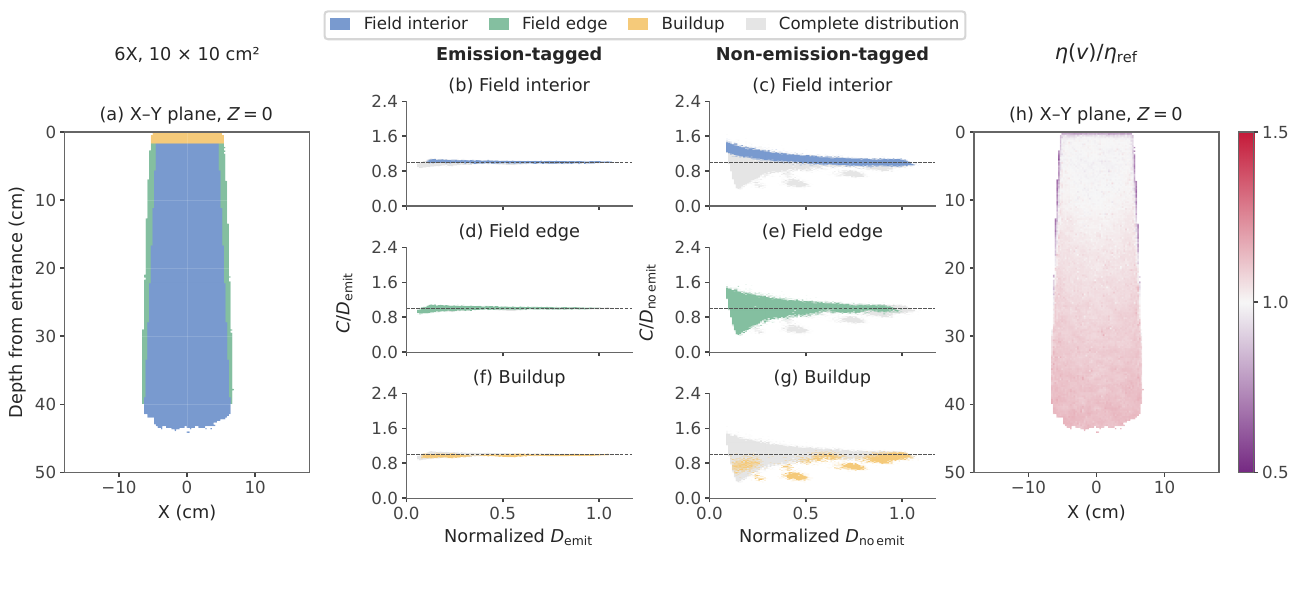}}
\captionsetup{font={small,stretch=1.0},skip=4pt}
\caption{Spatial origins of the emission-tagged and non-emission-tagged ratio distributions for 6X, \(10\times10\)~cm\(^{2}\).
(a) Spatial map of the analyzed phantom voxels, with colors identifying the field-interior, field-edge, and buildup regions. The central X--Y slice (\(Z=0\)) passes through the beam axis; X denotes lateral position, and Y is displayed as depth from the water entrance.
(b, d, f) Emission-tagged and (c, e, g) non-emission-tagged ratio distributions from Figure~\ref{fig:ratios} (first row, last two columns), colored by spatial region as in (a).
Gray shows the complete distribution for each component. (h) Map of \(\eta(v)/\eta_{\mathrm{ref}}\), the local emission-tagged dose fraction relative to that in the reference region, in the same central X--Y plane.
\label{fig:spatial_component_comparison}}
\end{figure}
\endgroup

In comparison, the emission-tagged ratio distribution does not show similarly pronounced branches. Figure~\ref{fig:spatial_component_comparison} shows that its dominant band near unity contains voxels from the field interior, field edges, and buildup region. The median emission-tagged ratio at the entrance surface changes from 0.497 before mass correction to 0.976 after correction (Figure~\ref{fig:spatial_component_comparison}f).
 
Away from the entrance surface, the photon yield per unit emission-tagged energy was largely independent of position, and the structure of the total-dose ratio came from where non-emission-tagged deposition was concentrated.

This study has several limitations.
The partition is defined at the level of Monte Carlo transport steps, so the component fractions and component ratios depend on the production threshold and step limits and cannot be measured.
Only the total-dose ratio is independent of the partition.
The components should therefore be read as model quantities that help explain where conversion factors vary.
We simulated only open, jaw-defined square fields in homogeneous water, and we analyzed photon creation without optical transport, tissue optics, or detection.
Heterogeneous media, modulated fields, and electron beams remain to be studied.

\section{Conclusions}

Using an existing digital linac Monte Carlo framework, we separated deposited energy according to actual Cherenkov production by the contributing transport step.
Across the simulated water-phantom conditions, emission-tagged dose exhibited a narrower dominant \(C/D\) band near unity  and closer normalized lateral profile agreement with local Cherenkov production than total or non-emission-tagged dose, as expected in part from its definition.
Departures of the total Cherenkov-to-dose ratio from unity followed the spatial variation of the emission-tagged fraction of dose, which varied most in the buildup region, at field edges, and at depth.
Analyses based on total dose combine emission-tagged and non-emission-tagged contributions, which can hide their different relationships with photon production.
Because the emission-tagged component cannot be measured directly and depends on transport settings, we see the decomposition mainly as a Monte Carlo tool for explaining where Cherenkov-based dose estimates need correction.

\section*{Author contributions}

Xianghuan He performed the simulations, analyzed the data, and drafted the manuscript. Yao Hao, Changyu Sun, Andy N. Wiles, Gregory B. Biedermann, Brian W. Pogue, and Rongxiao Zhang reviewed and revised the manuscript. All authors contributed to scientific discussions throughout the study.

\section*{Conflict of interest statement}

The authors have no relevant conflicts of interest to disclose.

\section*{Data and code availability}

The simulation data and custom simulation and analysis code supporting this study are available from the corresponding author upon reasonable request. Access to and redistribution of third-party phase-space files and treatment-planning inputs are subject to the relevant providers' conditions.

\section*{Figure captions}
\begingroup
\small

\noindent\textbf{Figure 1.} Voxelwise Cherenkov-to-dose ratios for \(3\times3\)~cm\(^{2}\) (left three columns) and \(10\times10\)~cm\(^{2}\) (right three columns) fields.
Rows show 6X, 6FFF, 10X, 10FFF, and 15X.
Within each field, columns correspond to total dose \(D_{\mathrm{total}}\), emission-tagged dose \(D_{\mathrm{emit}}\), and non-emission-tagged dose \(D_{\mathrm{no\,emit}}\).
The horizontal axis shows the indicated dose quantity \(D\), and the vertical axis shows \(C/D\), where \(C\) denotes mass-corrected local Cherenkov photon counts.
Each quantity is normalized to its own mean in the common reference region.
Colors show voxel counts per fixed two-dimensional bin on a logarithmic scale; dashed lines mark \(C/D=1\).
Only voxels with total dose of at least 10\% of its maximum are shown.\par

\noindent\textbf{Figure 2.} Lateral X profiles at \(Z=0\) for \(3\times3\)~cm\(^{2}\) (left two columns) and \(10\times10\)~cm\(^{2}\) (right two columns) fields.
Rows show 6X, 6FFF, 10X, 10FFF, and 15X; within each field, columns show \(d_{\mathrm{max}}\) and \(d_{10}\). 
Solid curves show total dose (\(D_{\mathrm{total}}\)), emission-tagged dose (\(D_{\mathrm{emit}}\)), non-emission-tagged dose (\(D_{\mathrm{no\,emit}}\)), and local Cherenkov photon counts (\(C\)), each normalized to its own mean in the common reference region.
Dashed curves show \(C-D\) for each dose quantity on the right axes; differences are dimensionless and the horizontal line denotes zero.\par

\noindent\textbf{Figure 3.} Spatial origins of the emission-tagged and non-emission-tagged ratio distributions for 6X, \(10\times10\)~cm\(^{2}\).
(a) Spatial map of the analyzed phantom voxels, with colors identifying the field-interior, field-edge, and buildup regions. The central X--Y slice (\(Z=0\)) passes through the beam axis; X denotes lateral position, and Y is displayed as depth from the water entrance.
(b, d, f) Emission-tagged and (c, e, g) non-emission-tagged ratio distributions from Figure~\ref{fig:ratios} (first row, last two columns), colored by spatial region as in (a).
Gray shows the complete distribution for each component. (h) Map of \(\eta(v)/\eta_{\mathrm{ref}}\), the local emission-tagged dose fraction relative to that in the reference region, in the same central X--Y plane.\par
\endgroup


\begin{thebibliography}{10}
\providecommand \doibase [0]{https://doi.org/}%

\bibitem{Glaser2014VideoRateCherenkov}
Glaser AK, Andreozzi JM, Davis SC, et al. Video-rate optical dosimetry and
  dynamic visualization of {IMRT} and {VMAT} treatment plans in water using
  {Cherenkov} radiation. {\it Med Phys.} 2014\string;41(6)\string:062102.
\newblock \href {\doibase 10.1118/1.4875704} {doi:10.1118/1.4875704}

\bibitem{Zhang2013SuperficialPhoton}
Zhang R, Glaser AK, Gladstone DJ, Fox CJ, Pogue BW. Superficial dosimetry
  imaging based on {\v{C}erenkov} emission for external beam radiotherapy with
  megavoltage x-ray beam. {\it Med Phys.} 2013\string;40(10)\string:101914.
\newblock \href {\doibase 10.1118/1.4821543} {doi:10.1118/1.4821543}

\bibitem{Glaser2014CherenkovDose}
Glaser AK, Zhang R, Gladstone DJ, Pogue BW. Optical dosimetry of radiotherapy
  beams using {Cherenkov} radiation: the relationship between light emission
  and dose. {\it Phys Med Biol.} 2014\string;59(14)\string:3789--3811.
\newblock \href {\doibase 10.1088/0031-9155/59/14/3789}
  {doi:10.1088/0031-9155/59/14/3789}

\bibitem{Zlateva2019Formalism}
Zlateva Y, Muir BR, El~Naqa I, Seuntjens JP. {Cherenkov} emission-based
  external radiotherapy dosimetry: {I}. Formalism and feasibility. {\it Med
  Phys.} 2019\string;46(5)\string:2370--2382.
\newblock \href {\doibase 10.1002/mp.13414} {doi:10.1002/mp.13414}

\bibitem{Zhang2017BeamTissueCherenkov}
Zhang R, Glaser AK, Andreozzi J, et al. Beam and tissue factors affecting
  {Cherenkov} image intensity for quantitative entrance and exit dosimetry on
  human tissue. {\it J Biophotonics.} 2017\string;10(5)\string:645--656.
\newblock \href {\doibase 10.1002/jbio.201500344} {doi:10.1002/jbio.201500344}

\bibitem{Agostinelli2003Geant4}
Agostinelli S, Allison J, Amako K, et al. {Geant4}---a simulation toolkit. {\it
  Nucl Instrum Methods Phys Res A.} 2003\string;506(3)\string:250--303.
\newblock \href {\doibase 10.1016/S0168-9002(03)01368-8}
  {doi:10.1016/S0168-9002(03)01368-8}

\bibitem{Allison2016Geant4}
Allison J, Amako K, Apostolakis J, et al. Recent developments in {Geant4}. {\it
  Nucl Instrum Methods Phys Res A.} 2016\string;835\string:186--225.
\newblock \href {\doibase 10.1016/j.nima.2016.06.125}
  {doi:10.1016/j.nima.2016.06.125}

\bibitem{Constantin2011TrueBeam}
Constantin M, Perl J, LoSasso T, et al. Modeling the {TrueBeam} linac using a
  {CAD} to {Geant4} geometry implementation: dose and {IAEA}-compliant phase
  space calculations. {\it Med Phys.} 2011\string;38(7)\string:4018--4024.
\newblock \href {\doibase 10.1118/1.3598439} {doi:10.1118/1.3598439}

\bibitem{Capote2006PhaseSpace}
Capote R, Jeraj R, Ma CM, et al. {\it Phase-space database for external beam
  radiotherapy: summary report of a consultants' meeting}.
\newblock INDC(NDS)-0484.
\newblock International Atomic Energy Agency; 2006.
\newblock Accessed October 19, 2026.
\newblock \url{https://www-nds.iaea.org/phsp/documents/indc-nds-0484.pdf}.

\bibitem{Cloutier2022Polarization}
Cloutier {\'E}, Beaulieu L, Archambault L. Direct in-water radiation dose
  measurements using {Cherenkov} emission corrected signals from polarization
  imaging for a clinical radiotherapy application. {\it Sci Rep.}
  2022\string;12(1)\string:9608.
\newblock \href {\doibase 10.1038/s41598-022-12672-w}
  {doi:10.1038/s41598-022-12672-w}

\bibitem{Glaser2013Projection}
Glaser AK, Davis SC, McClatchy DM, Zhang R, Pogue BW, Gladstone DJ. Projection
  imaging of photon beams by the {\v{C}erenkov} effect. {\it Med Phys.}
  2013\string;40(1)\string:012101.
\newblock \href {\doibase 10.1118/1.4770286} {doi:10.1118/1.4770286}

\end{thebibliography}
\end{document}